\documentclass{aa}
\usepackage{txfonts}
\usepackage{natbib}
\bibpunct{(}{)}{;}{a}{}{,} % to follow the A&A style
\usepackage{graphicx}

\title{Dust distribution during reionization}
\author{Erik Elfgren \inst{1}
        \and Fran\c cois-Xavier D\'esert \inst{2}
        \and Bruno Guiderdoni \inst{3}
}
\institute{Department of Physics, Lule\aa\ University of Technology, SE-971 87 Lule\aa, Sweden
                \and Laboratoire d'Astrophysique, Observatoire de Grenoble,
                     BP 53, 414 rue de la piscine, F-38041 Grenoble Cedex 9, France
		\and Universit\'e Lyon 1, Centre de Recherche, Astrophysique de Lyon, Observatoire de Lyon,
		     9 avenue Charles Andr\'e, F-69230 Saint Genis Laval ; CNRS, UMR 5574, France.
}
\offprints{Erik Elfgren,
\email{elf@ludd.luth.se}}

\date{Received <date> / Accepted <date>}

\begin{document}

\abstract
{ The dust produced by the first generation of stars will be a foreground to cosmic
microwave background.}
{ In order to evaluate the effect of this early dust, we calculate
the power spectrum of the dust emission anisotropies
%and show that this dust might be detectable with the Planck satellite
%at small angular scales ($\ell \gtrsim 1000$).
and compare it with the sensitivity limit of the Planck satellite.}
{ The spatial distribution of the dust is estimated through the distribution
of dark matter.}
{ At small angular scales ($\ell \gtrsim 1000$) the dust signal is
found to be noticeable with the Planck detector for certain values of dust lifetime and
production rates. The dust signal is also compared to sensitivities of other instruments.
The early dust emission anisotropies are finally compared to those of local dust
and they are found to be similar in magnitude at mm wavelengths.
}
{}

%At mm wavelength, early dust could be at a similary to 
%contamination level, foreground

%

\maketitle

%--------------------------------------------------------------------
\section{Introduction}
%--------------------------------------------------------------------

The importance of the cosmic microwave background (CMB) as a cosmological tool
has been demonstrated thoroughly during the last few years.
It has been used to evaluate the age, the Hubble
parameter, the baryon content, the flatness and the optical depth of the
reionization of the universe \citep{2003ApJS..148....1B}.
It has also been used to set upper limits on 
the non-Gaussianity of the primary fluctuations \citep{2003ApJS..148..119K},
the Sunyaev-Zeldovich fluctuations from the first stars \citep{2003MNRAS.342L..20O},
the primordial magnetic fields \citep{2003MNRAS.344L..31S},
the spatial curvature of the universe \citep{2003MNRAS.343L..95E},
the formation of population III stars \citep{2003ApJ...591L...5C},
and the neutrino masses using only WMAP data \citep{2005PhRvD..71d3001I}
as well as combining them with other data \citep{2003JCAP...05..004H}.
%and ditto with only WMAP data \citep{2005PhRvD..71d3001I}.
%Measurements of the CMB can give is vital information about the lifetim...
However, in order to interpret the CMB signal correctly, its foregrounds
must also be well known.

In this paper we focus on one particular aspect of the foreground of the CMB:
the dust from the first generation of stars. It is here assumed that dust was
created during the reionization period in the first generation of stars and
was then ejected into the intergalactic medium (IGM).
The dust is heated by the ionizing photons to a temperature slightly above
$T_{CMB}$.
The net effect on the CMB is a small monopole distortion of the CMB 
with a characteristic electromagnetic spectrum close to the CMB primary
anisotropy ($\Delta T$) spectrum times the frequency squared.
This effect was studied in \cite{2004A&A...425....9E}.

%The net effect on the CMB is 1) a monopole distortion of the CMB which
%is small, owing to the energy density of ionizing photons relative to the CMB
%2) secondary anisotropies with a characteristic spectrum close to the CMB primary
%anisotropy ($\Delta T$) spectrum times the frequency squared.
%The effect of the dust on the CMB has been studied in \cite{2004A&A...425....9E}.

%The dust partly absorbs the CMB photons and
%slightly distorts the CMB spectrum. As we have shown in an earlier paper
%\cite{2004A&A...425....9E}, this dust has a characteristic spectrum
%proportional to a primary CMB anisotropy ($\Delta T$) spectrum times the
%frequency squared. The dust spectrum is lower than the CMB by roughly two
%orders of magnitude because the heating from the stars is significantly less
%than that of the CMB at the time.

Moreover, the dust also has a characteristic spatial distribution,
which could be used to identify its signal.
The distribution gives rise to anisotropies in the dust emission,
which can be measured with several current and future experiments.
%TODO as measured by the anisotropy experiments
%The objective of this paper is to determine this distribution and its
%impact on different measurements of the CMB.
The objective of this paper is to determine this spatial distribution
and its resulting anisotropies. %, in order to compare with current and 
%future experiments. 
%as measured by the anisotropy experiments.
Of particular interest is the
Planck satellite mission, but other instruments are also useful, like %ALMA, SCUBA, MAMBO and BLAST could be of interest.
ALMA \citep{2003SPIE.4837..110W},
%FIRAS II, \cite{2002ApJ...581..817F},
BLAST \citep{2001dmsi.conf...59D},
% cf also 2004SPIE.5498...42D - but only abstract
BOLOCAM (LMT) and (CSO) \citep{2000ASPC..217..115M},
MAMBO \citep{2004MNRAS.354..779G},
SCUBA \citep{1999MNRAS.308..527B}, and
SCUBA2 \citep{2004SPIE.5498...63A}.

The spatial distribution of the dust is estimated with the help of GalICS (Galaxies In
Cosmological Simulations) $N$-body simulations of dark matter (DM) \citep{2003MNRAS.343...75H},
which are described in more detail in Sect.~\ref{sec:DM}. The dust distribution is then
combined with the intensity of the dust emission, and this is integrated along
the line of sight. The resulting angular power spectrum is then computed as
$C_\ell$ and compared with the detection limits of Planck.

In the following, we assume a $\Lambda$CDM universe with
$\Omega_{tot} = \Omega_m + \Omega_\Lambda = 1$, where
$\Omega_m = \Omega_b + \Omega_{DM} = 0.133/h^2$, $\Omega_b = 0.0226/h^2$,
$h = 0.72$ and $\tau_e=0.12$, as advocated by WMAP \citep{2003ApJS..148..175S},
using WMAP data in combination with other CMB datasets and large-scale structure observations %Myref 80p15t7.
(2dFGRS + Lyman $\alpha$).

%Epoch of reionization, paramters of the univ. TOC?

%--------------------------------------------------------------------
\section{Dark matter simulation}
%--------------------------------------------------------------------
\label{sec:DM}
% TODO: figures? radial/transverse replication, more from GalICS paper?

The distribution of DM in the universe is calculated with the GalICS
 program.
The cosmological N-body simulation we refer to throughout this paper is done
with the parallel tree-code developed by \cite{1999ninin}. The power spectrum
is set in agreement with \cite{1996MNRAS.282..263E}: $\sigma_8$ = 0.88, and
the DM-density field was calculated from $z=35.59$ to $z=0$, outputting 100
snapshots spaced logarithmically in the expansion factor.

%In order to evaluate the spatial distribution of the dust, we use a simple model

The basic principle of the simulation is to randomly distribute a number of
DM-particles $N^3$ with mass $M_{{\rm DM}}$ in a box of size $L^3$.
Then, as time passes, the particles interact gravitationally, clumping together
and forming structures.
%The clumps of Dark Matter are called halos and in our
%simulation we require at least 5 particles to clump together before we call it
%a halo.
When there are at least 5 particles together, we call it a DM-clump.
There are supposed to be no forces present other than gravitation,
and the boundary conditions are assumed to be periodic.

In the simulation we set the side of the box of the simulation to
$L=100h^{-1}$ Mpc and the number of particles to $256^3$, which implies a
particle mass of $\sim 5.51\times 10^{9}h^{-1} M_\odot$. Furthermore, for the
% rho_c*L^3/M_sun/N_part*Omega_DN =
% 1.9e-26 * h^2 * (100 * 3.08568025e22/h)^3/1.989E30/256^3 * 1/3 = 5.58e9
simulation of DM, the cosmological parameters were
$\Omega_\Lambda = 2/3$, $\Omega_m = 1/3$ and $h=2/3$.
The simulation of the DM was done before the results from WMAP
were published, which
explains the difference between these parameters and the values %adopted in the
used elsewhere in this paper, as stated in the introduction.
%end of the introduction.
Fortunately, the temporal distribution of the dust is independent of the
value of $h$, which means that the impact of this small discrepancy is
not important.
Between the assumed initial dust formation at $z\sim 15$ and the end of
this epoch in the universe at $z\sim 5$, there are 51 snapshots. In each snapshot
a friend-of-friend algorithm was used to identify virialized groups of at least
5 DM-particles. 
%This number of required particles means that the first halos in our model
%formed at $z=14.7$.
%The number of particles have been set low in order to
%produce halos already at $z=14.7$.
For high resolutions, it is clear that the mass resolution
is insufficient. Fortunately, the first 5-particle DM-clump appears at $z=14.7$, while
the bulk of the luminosity contribution comes from $z\lesssim 12.5$.
At $z=12.6$ there are 19 clumps and at $z=12.2$ there are 45 clumps (and rapidly
increasing in number) so it should be statistically sufficient.

In order to make a correct large-scale prediction of the distribution of the
DM and therefore the dust, the size of the box would have to be of
Hubble size, i.e., $\sim3000h^{-1}$ Mpc.
However, for a given simulation time, increasing the size of the box and
maintaining the same number of particles would mean that we lose out in mass
resolution, which is not acceptable if we want to reproduce a fairly realistic
scenario of the evolution of the universe.

There is another way to achieve the desired size of the simulation without
losing out in detail or making huge simulations. This method is called MoMaF (Mock
Map Facility) and is described in detail in \cite{2005MNRAS.360..159B}. The
basic principle is to use the same box, but at different stages in time and
thus a cone of the line of sight can be established. In order to avoid
replication effects, the periodic box is randomly rotated for each time-step.
This means that there will be a loss of correlation information on the edges of
the box, since these parts will be gravitationally disconnected from the
adjacent box. Fortunately, this loss will only be of the order of 10\%, as shown
%in \cite{2003astro.ph..9305B}.
in \cite{2005MNRAS.360..159B}.
For scales larger than the size of the box, there is obviously no  
information whatsoever on correlation from the simulation.

\subsection{Validity of the simulation}
%This simulation has been tested, see \cite{GalICS}.
GalICS is a hybrid model for hierarchical galaxy formation studies, combining
the outputs of large cosmological N-body simulations with simple, semi-analytic
recipes to describe the fate of the baryons within DM-halos. The
simulations produce a detailed merging tree for the DM-halos,
including complete knowledge of the statistical properties arising from the
gravitational forces.

The distribution of galaxies resulting from this GalICS simulation has been
compared with the 2dS \citep{2001MNRAS.328.1039C} and the Sloan Digital Sky
Survey \citep{2001misk.conf..249S} and found to be realistic on the angular
%scales of $3' \lesssim \theta \lesssim 30'$, see \cite{2003astro.ph..9305B}.
scales of $3' \lesssim \theta \lesssim 30'$, see \cite{2006MNRAS.369.1009B}.
The discrepancy in the spatial correlation function for other values of
$\theta$ can be explained by the limits of the numerical simulation. Obviously,
any information on scales larger than the size of the box ($\sim 45$') is not reliable.
%{\it More volume effects discussion?}
%Correlations on angular scales larger than 45' are not be available from our simulations.
Fortunately, the dust correlations increase at smaller angles, while the CMB and many other
signals decrease. This means that our lack of information on angular 
scales $\theta>45$' ($\ell\lesssim 250$) will not be important, as can be seen in Fig.~\ref{fig:f353}.
The model has also proven to give reasonable results for Lyman break galaxies at $z=3$
\citep{2004MNRAS.352..571B}.
It is also possible to model active galactic nuclei using the same model \citep{2005MNRAS.364..407C}.

%Since the simulation gives reasonable
%predictions of the matter distributions today, it seems likely that it is also
%valid at higher $z$ when the early dust is produced.
Since it is possible to reproduce reasonable  
correlations from semi-analytic modeling of galaxy formation within  
this simulation at $z=0-3$, we hereafter attempt to do so at higher $z$ values, when  
the early dust is produced.

%--------------------------------------------------------------------
\section{Model}
%--------------------------------------------------------------------
\label{sec:Model}
Since very little is known about the actual distribution of the dust
throughout the universe at this time, we simply assume that the dust
distribution follows the DM-distribution.
We propose and explore two
different ways for the dust distribution to follow the DM-distribution. The first is
to let the dust be proportional to the DM-clumps, the second
is to make a hydrodynamical smoothing of the DM-density
field and set the dust density proportional to this density.
In both cases we assume that
\begin{equation}
         \rho_{dust}(\vec r, z) \propto \rho_{DM}(\vec r, z),
\end{equation}
where $\rho_{DM}$ represents either the clump method or the smoothing
method density.
The stars that produce the dust are probably formed close to
gravitational hot-spots in the universe and these spots are shaped
by the DM. This means that the production of dust can be well
approximated by the DM distribution.
According to \cite{2006ApJ...640...31V}, the initial velocity of the dust is in the order of $10^5$ m/s
and the DM-simulations give halos with an escape velocity that is $v_e\gtrsim 10^5$ m/s, given that
their mass is $\gtrsim 4.1\times 10^{10}$ M$_\odot$.
This means that most of the dust will stay near the halo and
% sqrt(2*G*0.2741509E+01*1e11*M_sun/(0.3875238E-01*Mpc))
%The subsequent redistribution
%of dust due to their inital velocity 
we will therefore focus on the clump method.
% since the dust was created in galaxies
%that are found in the DM-clumps. The dust will propagate to some extent
%but, for simplicity, we assume that it stays close to the DM-clump.
The hydrodynamical smoothing case is included for reference only.

%We will focus on the Halo method since it is more likely
%that the dust will have formed in galaxies and halos than that
%it will have formed anywhere where there is dark matter.

\begin{figure}%[here!]
        \resizebox{\hsize}{!}{\includegraphics{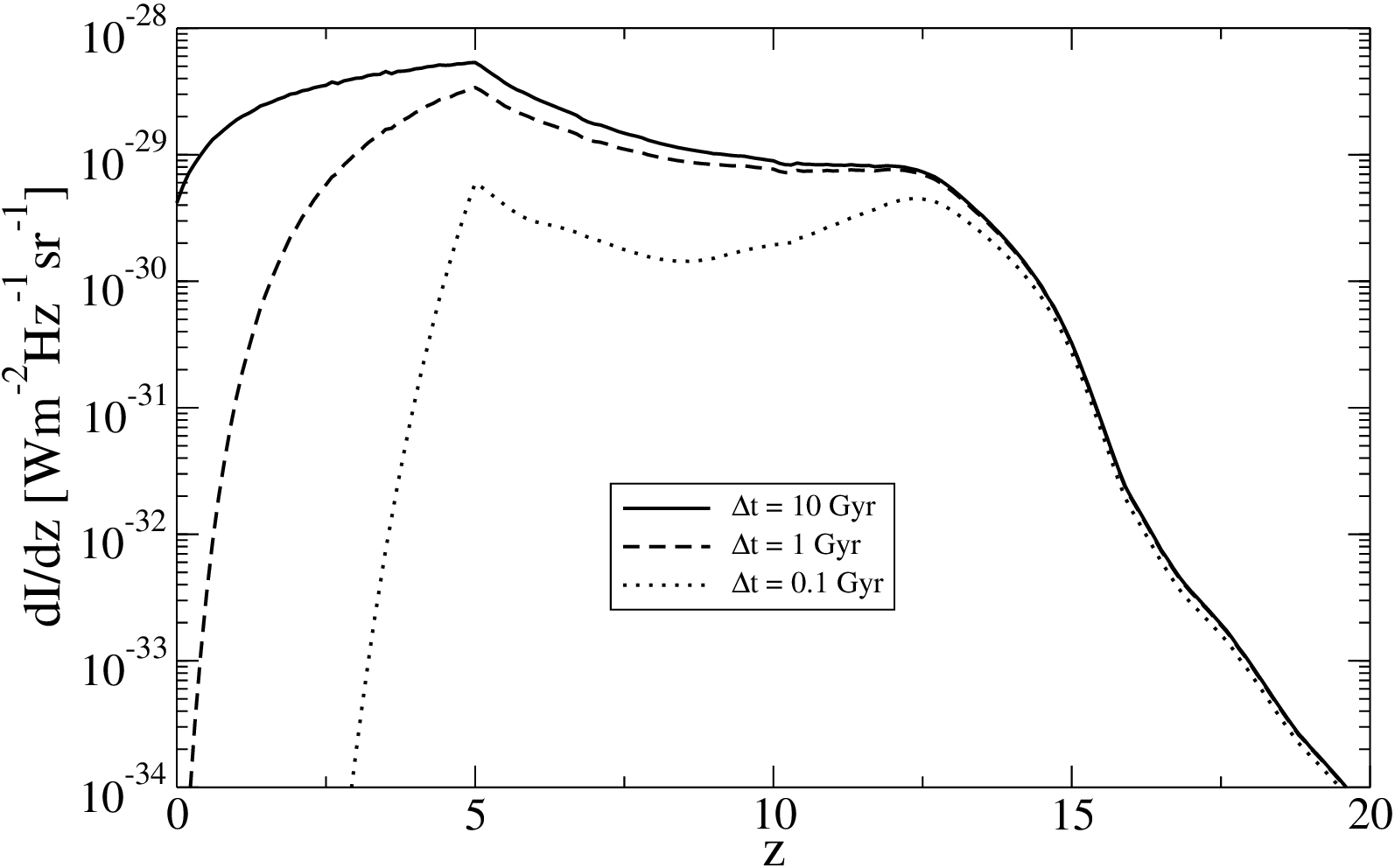}} % dIdz
        \caption{Intensity contribution from the dust per time-step $z$
		integrated over all frequencies. 
		This model assumes that the fraction of metals produced in stars
		that end up as dust is $f_d=0.3$. The mean dust lifetime is a largly
		unknown parameter, and therefore three different values are explored,
		$\Delta t=0.1$, 1 and 10 Gyr.}
        \label{fig:dIdz}
\end{figure}

In order to estimate the measured intensity, we need to calculate this
distribution in terms of the intensity from the dust emission.
%{\bf NO In our previous paper, \cite{2004A&A...425....9E}, we calculated
%the intensity as a function of redshift, supposing that none
%of the light emitted from the stars is absorbed. This is close
%enough to the truth since there are more than 100 ionizing
%photons produced per baryon. }
The early dust is optically thin,
and its intensity as a function of redshift has been calculated in
\cite{2004A&A...425....9E} and is shown in Fig.~\ref{fig:dIdz}.
This model assumes that the fraction of metals produced in stars
that end up as dust is $f_d=0.3$. The mean dust lifetime is a largly
unknown parameter and therefore three different values are explored,
$\Delta t=0.1$, 1, 10 Gyr.
%In the present This intensity from the dust is distributed such that 

In our present model, we put the spatial distribution of the dust intensity to
\begin{equation}
        \frac{dI}{dz}(\vec r, z) = \frac{dI(z)}{dz} \cdot \frac{\rho_{DM}(\vec r, z)}{\langle\rho_{DM}\rangle(z)},
\label{eq:Irho}
\end{equation}
where $dI(z)/dz$ is the dust intensity at redshift $z$ as measured at $z=0$
and $\langle\rho_{DM}\rangle(z)$ is
the mean DM-density at redshift $z$.
The MoMaF method (see Sect.~\ref{sec:DM}) is then used to project the emitted
intensity from the dust on a $45'\times45'$ patch along the line of sight.
%Within each box equation \ref{eq:Irho} is normalized such that
%\begin{equation}
%       \int_{\textrm{box depth}}\frac{dI}{dz}(z) = \sum_{i_x=1}^{N} \sum_{i_y=1}^{N}
%               I_{i_x,i_y}/N^2
%\end{equation}
%where $I_{i_x,i_y}$ is the intensity from grid point $(i_x, i_y)$
%and $N^2$ is the total number of grid points.
The contribution from each simulated box is added, and the integrated
dust intensity is calculated.

For $z>2.3$, the time-steps are smaller than the size of the box
and each box overlaps with the next box along the line of sight. 
However, for $z<2.3$ the time-steps were simulated too far apart
and when we pile the boxes, there will be a small part of the
line of sight that will not be covered.
%If a box is deeper than the time-step is long, the box is simply
%chopped at $c\cdot t_s$, where $t_s$ is the time-step.
%For $z<2.3$, the time-step is longer than the box is deep, which
%means that there will be gaps in the Dark Matter density distribution.
Fortunately, this is of little consequence since the dust intensity is low at this time,
and the gap is small. 
Each box ($45'\times45'$) is divided into a grid according to the resolution that we
wish to test. For Planck this means a grid that is 18$\times$18 pixels$^2$,
for SCUBA 45$\times$45 pixels$^2$.

%Our simulated box of 45'$^2$ will thus
%correspond to 18$\times$18 pixels in Planck. Correlations on larger angular scales
%than 45' will not be available from our simulations. However, the
%dust correlations will increase at smaller angles while the CMB and many other
%signals will decrease. This means that our lack of information on angular
%scales $\ell\lesssim 250$ will not be of any consequence, as can be seen in
%Fig. \ref{fig:f353}.
%TODO earlier?

%{\bf 18*18 would have been better} 

To check the normalization of the resulting intensity image, we have calculated its $\sum dI_{x,y}/N_{pix}^2$,
where $dI_{x,y}$ is the observed intensity on pixel $(x, y)$ and $N_{pix}^2$ is the number of pixels$^2$,
and found it to be equal to $\int dI(z) dz$ to within a few per cent.

%{\it Method}\\
%{\it I've checked by taking the mean of the resulting image -> dI.
%{\it $\sum dI\cdot dz = 0.1\sum dI_i \approx 8.74$e-13 ($\Delta t = 0.1$ Gyr)
%       $\sum dI_{x,y} \approx 8.65$e-13 -- should be /81 -> *81.} \\

%{\it Figure of Cl with only DM?}
%--------------------------------------------------------------------
\section{Results and discussion}
%--------------------------------------------------------------------
\label{sec:Results}

\begin{figure}%[here!]
        \resizebox{\hsize}{!}{\includegraphics{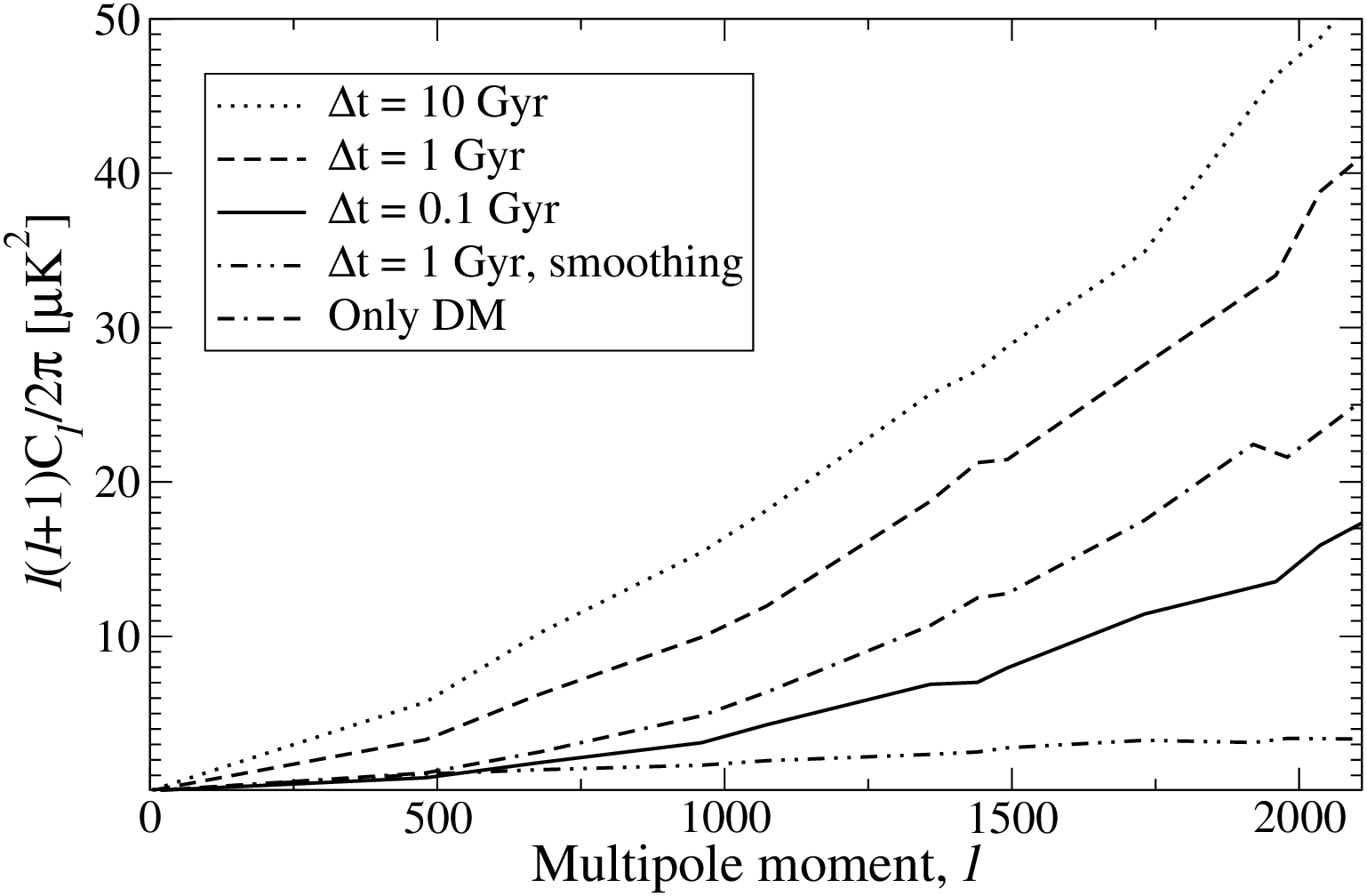}} % Cl9.pdf
        \caption{Dust power spectrum in CMB thermodynamic units 
          at 353 GHz for a map $45'\times 45'$ and Planck resolution $5'$
          for three different lifetimes for the dust particles, 0.1,
          1, and 10 Gyrs, with a solid, dashed and dotted line,
          respectively.
	%The dash-dash-dot line respresents the
         %normalized correlation for the DM halos only without dust.
          The DM smoothing method for a dust lifetime of 1 Gyr is the
          dot-dot-dash line.
	% We note that the DM smoothing method
          %gives correlations that are approximately a factor 10 lower
          %than the clump method. Also, the form is not quite the
          %same.
                %The dotted line was obtained by pasting the 400 images in a 20$\times$20
                %grid and calculating the $C_\ell$ on this image. The dashed line represents
                %the hydrodynamical smoothing.
}
        \label{fig:Cl9}
\end{figure}

As described above, the MoMaF technique produces an image of the line of sight.
This image represents the patch of the sky covered by the box, 150 co-moving Mpc$^2$,
which translates to $\sim 45$ arcmin$^2$ at $z=14.7$.
In order to avoid artifacts at the edges, the image is apodized,
whereafter it is Fourier transformed
into frequency space $P_k$. In order to convert this spectrum into spherical-harmonics
correlation functions we apply the following transformation:

\begin{equation}
	\ell = k 2\pi / \theta,
\end{equation}
\begin{equation}
	C_\ell = \theta^2 C_k,
\end{equation}
%\begin{eqnarray}
%        \ell = k 2\pi / \theta,\\
%        C_\ell = \theta^2 C_k,
%        %\ell &=& k \cdot (2 \pi)/(4\cdot \sin(\theta/4))\cdot N_{map}/N_{fft} \\
%        %C_\ell &=& \theta^2 C_k.
%\end{eqnarray}
where $\theta$ is the size in radians of the analyzed box.
These $C_\ell$ are then calculated at a frequency $\nu = 353$ GHz,
which is one of the nine Planck frequency channels. %and is central in
%the following discussion.
%{\bf I do not understand this unit, suppress in units of  in units of [$\mu$K$^2/B_\nu(T_{CMB})$] }
%(one of the Planck detectors work at this frequency)
As found in \cite{2004A&A...425....9E},
the intensity is proportional to the frequency squared, which means that the
power spectrum
%(expressed in CMB thermodynamic units, $\mu$K$^2_\textrm{CMB}$)
from the dust at a frequency $\nu$ is
\begin{equation}
        C_\ell(\nu) = C_\ell(353\textrm{ GHz})\cdot\left(\frac{\nu}{353\textrm{ GHz}}\right)^4,
\label{eq:freqdep}
\end{equation}
where $C_\ell$ is given in terms of $\mu$K$^2_\textrm{CMB}$.
In order to estimate an average power spectrum, 400 such images were generated
and the $C_\ell$ were averaged over these. For comparison, we also tried to paste
all these images together and calculate the $C_\ell$ for this (180$\times$180 pixels$^2$) image.
The result was similar to the average $C_\ell$.
To validate our results, we also calculated the variance of the images
and compared them with $\sum_{\ell}\frac{2\ell+1}{4\pi}C_\ell$, and found them
to be compatible.
%We have also computed the $C_\ell$ for an image consisting
%of 20$\times$20 ordinary images and found that them to be similar to
%the normal method.
The resulting power spectra can be seen in Fig.~\ref{fig:Cl9}.
The lifetime of these dust particles is a largely unknown factor, and we plot three different
lifetimes, 0.1, 1 and 10 Gyrs
\cite[for a more detailed discussion of dust lifetimes, see][]{1990eism.conf..193D}.
Furthermore, the dust intensity is proportional to the fraction of
the formed metals that actually end up as dust, which we have assumed to be $f_d=0.3$.
This means that the dust power spectrum is
\begin{equation}
C_\ell(f_d) = C_\ell(f_d=0.3)\cdot (f_d/0.3)^2.
\end{equation}
We note that there is only a small difference between dust lifetimes of 10 Gyrs and 1 Gyr, while the
one at 0.1 Gyr is lower by a factor of four.
The lowest curve in the figure represents the hydrodynamical smoothing method of distributing
the dust for a dust lifetime of 1 Gyr. Naturally, it is lower than the corresponding
$C_\ell$ for the clump method, since the DM-clumps are much more grainy (especially early in history)
than the smoothed DM field. The power spectra of the two methods differ by a factor of $\sim 10$ but they
do not have exactly the same form.

\begin{figure}%[here!] % DeltaForm.eps
        \resizebox{\hsize}{!}{\includegraphics{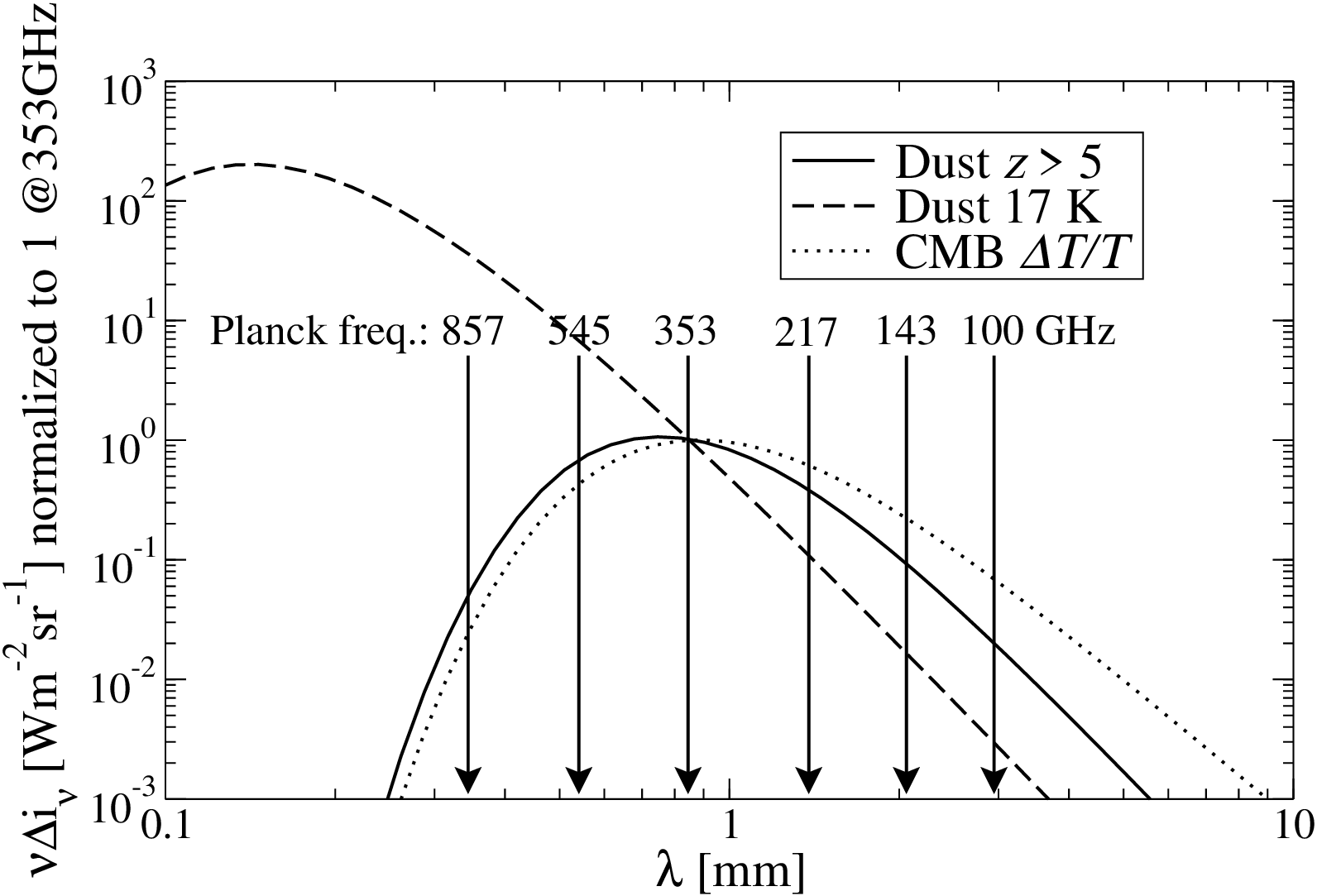}}
        \caption{The form of the early dust spectrum compared to the form
                of galactic dust (with a temperature of 17 K) and the CMB
		along with indicators of the Planck HFI frequencies.
                The curves have been normalized to unity at 353 GHz.
                We see that the early dust has a special spectral
                signature.
%{\bf Show Planck frequency channels}
                }
        \label{fig:Form}
\end{figure}

%\subsection{Dust Spectrum}
%{\it Use subsection Dust Spectrum?}\\
The dust frequency spectrum is distinctly different from that of other sources
in the same frequency range.  In Fig.~\ref{fig:Form}, we compare this
spectrum with that of the primoridal CMB $\Delta T$  anisotropies and
that of galactic dust, $T=17$ K \citep{1996A&A...312..256B}. In order to focus
on the forms of the spectra, we normalize the three curves to one at $\nu = 353$ GHz.
In case of a weak early dust signal,
this frequency signature could help us identify the signal by component separation
spectral methods.
%this frequency signature
%allow us to distinguish the dust signal from the
%CMB and other foregrounds.

\subsection{Detectability with Planck}
The Planck satellite\footnote{
	Homepage: http://www.rssd.esa.int/index.php?project=Planck
}, due for launch in 2007, will have an angular resolution of
$30' - 5'$ and will cover the whole sky.
The Planck high frequency instrument (HFI)
will measure the submillimeter sky at $\nu = $ 100, 143, 217, 353, 545,
and 857 GHz. We have chosen $\nu = 353$ GHz as our reference frequency. At higher
frequencies, the galactic dust will become more of a problem, and at lower
frequencies the CMB primary anisotropies will dominate.
% \lambda = 2.996 2.095 1.381   ==0.849==   0.550 0.350 mm

In order to test the detectability of the dust with Planck, we evaluate
the sensitivity,
%$C_\ell$:
%$\ell(\ell+1)C_l/2\pi$:
\begin{equation}
        \sigma = \sigma^{CMB}_\ell + \sigma^{instr}_\ell = \sqrt{\frac{2}{(2\ell+1) f_{cut} L}} \times (E_{CMB} + E_{instr})
\label{eq:ErrTot}
\end{equation}
\citep{1997PhRvD..56.4514T} % Eq 12
where $f_{cut}=0.8$ is the fraction of the sky used,
$L$ is the bin-size, $E_{CMB} = \ell(\ell+1)C_\ell^{CMB}/2\pi$ \citep{Lambda} is the cosmic variance. The instrument error
%over $\ell(\ell+1)C_l/2\pi$
is % Eq 1-3
\begin{equation}
        E_{instr} = C_\ell^{instr} \frac{\ell(\ell+1)}{2\pi} = f_{sky}\frac{4\pi s_X^2}{t_{obs}} \cdot e^{\ell^2\cdot\sigma_b^2} \cdot \frac{\ell(\ell+1)}{2\pi},
\label{eq:Eerr}
\end{equation}
where $f_{sky}=1$ is the fraction of the sky covered,
$s_X$ is the noise [$\mu$Ks$^{1/2}$],
$t_{obs}=14\cdot 30\cdot 24\cdot 3600$ s is the observation time (14 months), and
$\sigma_b=FWHM/2.35$ (FWHM is the full width at half maximum of the
beam in radians).
For Planck, the values of these parameters are given in Table~\ref{tab:PlanckParams}.
% which was obtained from \cite{2004AdSpR..34..491T}.
\begin{table}
%\begin{tabular}{|l||l|l|l|l|l|l|}
\caption{Parameters of the PLANCK HFI detector properties \citep{bluebook}}
\begin{tabular}{lllllll}
\hline
\hline
        Frequency [GHz] & 100 & 143 & 217 & 353 & 545 & 857 \\
%\hline
\hline
        $FWHM$ [$'$]      & 9.5 & 7.1 & 5.0 & 5.0 & 5.0 & 5.0 \\
%\hline
        $s_X$ [$\mu$Ks$^{1/2}$] & 32.0& 21.0& 32.3& 99.0& 990 & 45125 \\
\hline
\end{tabular}
\label{tab:PlanckParams}
\end{table}
%The values of the cosmic variance $E_{CMB}$ is the $\ell(\ell+1)C_\ell^{CMB}/2\pi$ itself (\cite{Lambda}).
%$\Lambda$-website: http://lambda.gsfc.nasa.gov/.

\begin{figure}%[here!]
        \resizebox{\hsize}{!}{\includegraphics{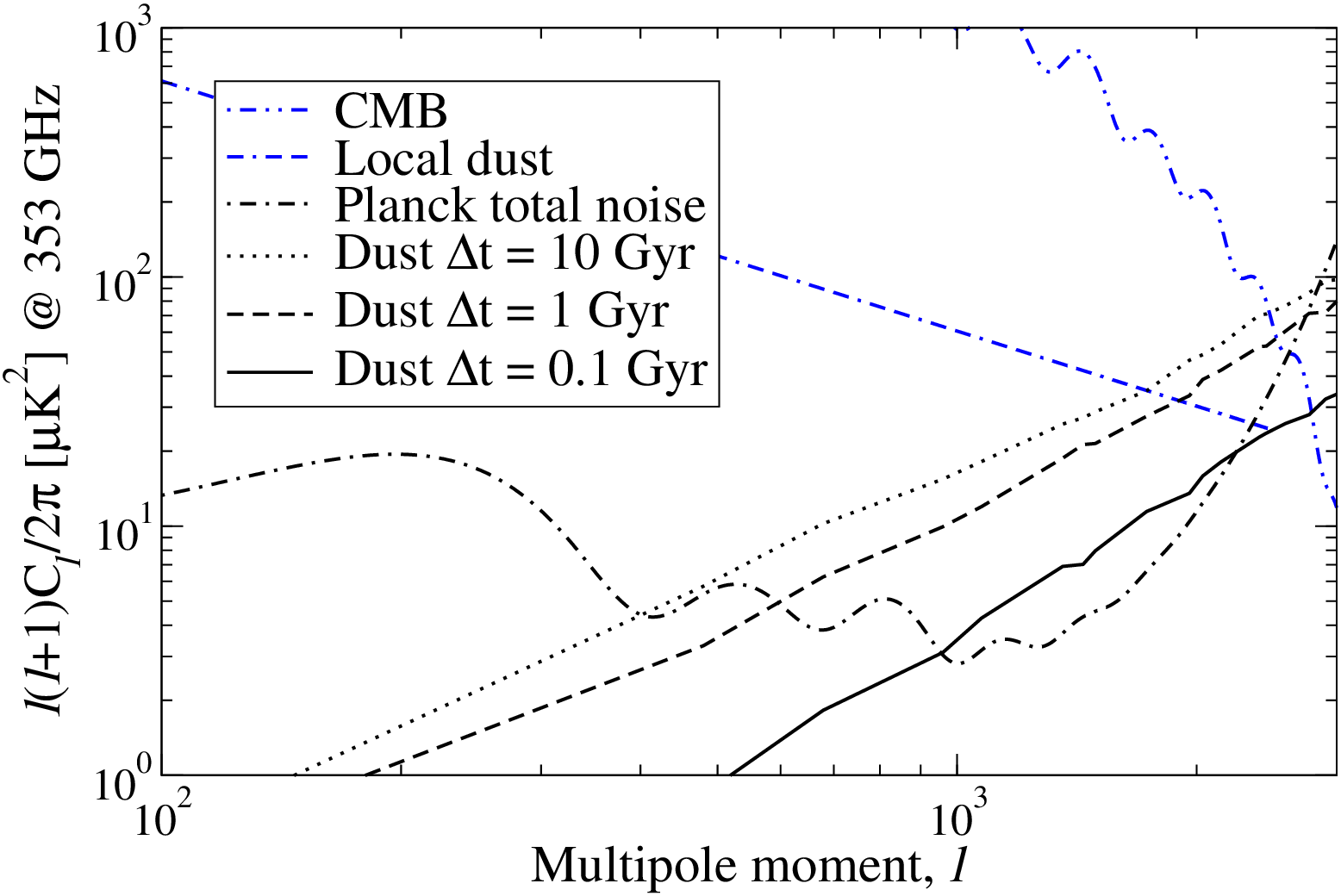}} % f353localdust.eps
        \caption{Comparison between dust power spectrum, the Planck error limits and local dust ($T=17$ K) at 353 GHz with binning 500.
                The error limits (total noise) consist of two parts; the CMB cosmic variance, which dominates for small $\ell$,
                and the instrument noise, which dominates for high $\ell$.}
        \label{fig:f353}
\end{figure}

\begin{figure*}%[here!]
	\sidecaption
	\centering
        \includegraphics[width=12cm]{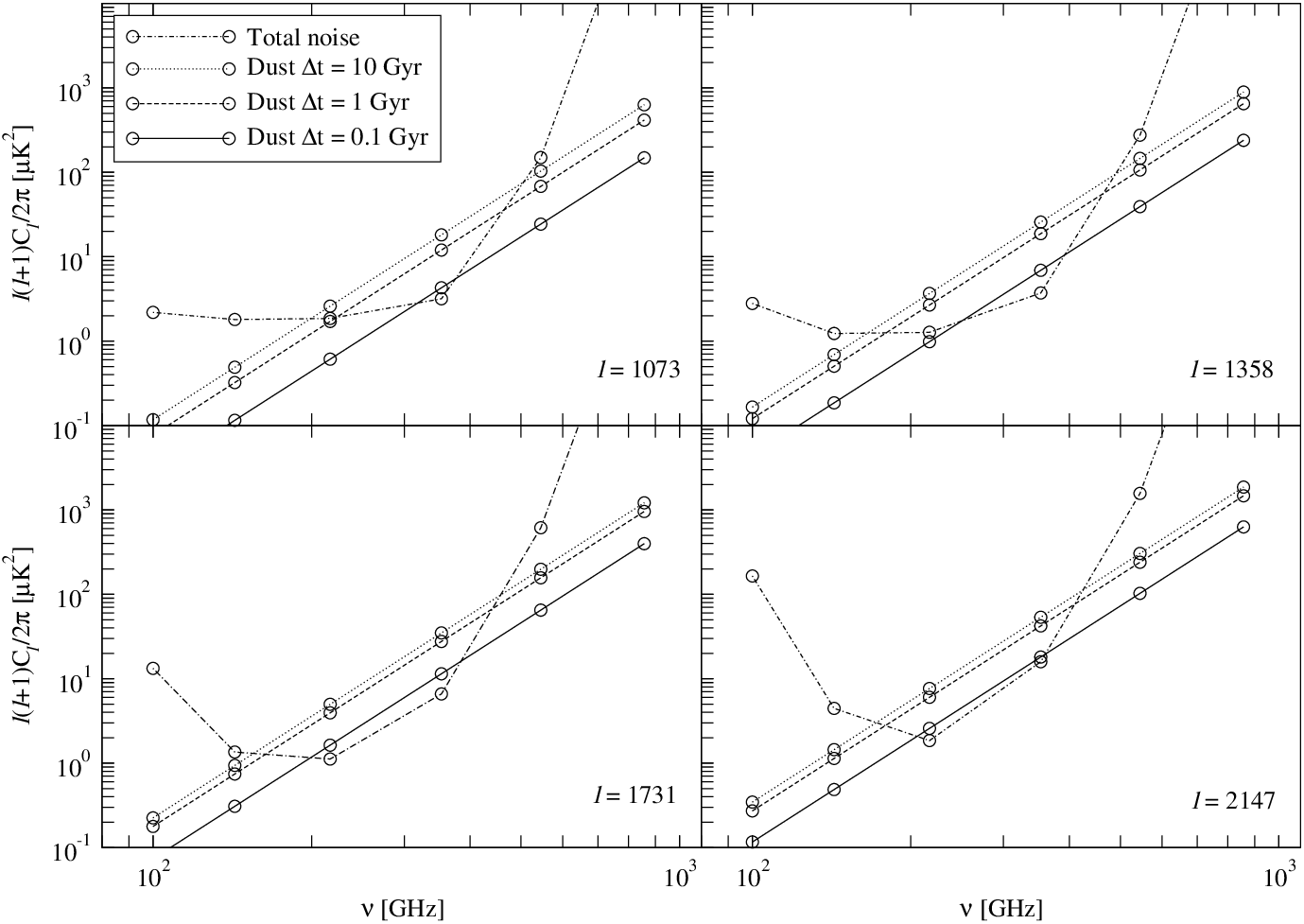} % AllL.pdf
        \caption{Comparison between dust power spectrum and Planck error limits at $\ell=1073$, 1358, 1731, and 2110 with binning 500.
        The error limits (total noise) consist of two parts; the CMB cosmic variance, which is constant at such $\mu$K$^2$-levels,
	%($\approx 1.0$)
        and the instrument noise, which has the shape of an exponential$\times\ell^2$.}
        \label{fig:allfL}
\end{figure*}

The resulting errors
for a binning of $L=500$ along with the dust power spectrum
is plotted in Figs.~\ref{fig:f353}-\ref{fig:allfL}.
In Fig.~\ref{fig:f353}, the frequency $\nu = 353$ GHz is fixed,
while $\ell$ is varied.
We note that $\ell \sim 1000$ seems to be a good value to search for dust.
At low $\ell$, the error due to the cosmic variance dominates, whereas at high $\ell$
the instrument noise dominates.

In Fig.~\ref{fig:allfL}, the $\ell$
multipole binning center
is fixed, 
and we show the electromagnetic spectrum of the primordial anisotropies
and the early dust emission.
%while the Planck frequencies constitute the variable.
The fourth
point from the left in the figures corresponds to $\nu = 353$ GHz and gives the best
signal over noise ratio. At low $\ell$ the cosmic variance is important,
at high $\ell$, the instrument noise dominates.

Early dust may therefore produce a measurable disturbance in the
primordial anisotropy angular power spectrum at high multipoles (in
the Silk's damping wing). Although the primary contaminant to the CMB
in the submillimeter domain is the interstellar dust emission, this
new component vindicates the use of more than two frequencies to
disentangle CMB anisotropies from submillimeter foregrounds.

Component separation methods for the full range of Planck
frequencies should be able to disentangle the CMB anisotropies from early
dust, far-infrared background fluctuations, and galactic dust
emission.
The Sunyaev-Zel'dovich (SZ) effect also rises in the submillimeter range
but the two central frequencies of the dust,
217 and 353 GHz, can be used for its identification since the SZ-effect drops more sharply at higher wave-lengths
(the SZ-effect is practically null at 217 GHz, quite different from the early dust spectrum).
The spectral shape of other foregrounds can be found in \citet[Fig.~1]{2000A&A...357....1A}
where we can also see that the region around $\ell\sim 2000$ is the most favorable for dust
detection.
%We find that the SZ is nearly 0 at 217 GHz whereas it is a third of the 353 value for early dust.
%    100            0.08      -4.07
%    143            0.16      -2.76
%    217            0.38      -0.03
%    353            1.00       5.83
%    545            2.38      13.97
%    857            5.89      26.85
%The power spectrum Fig. X.

\subsection{Detectability with other instruments}
%\subsection{Discussion}

There are several other instruments that might be used to detect the early dust:
ALMA \citep{2003SPIE.4837..110W},
%FIRAS II \citep{2002ApJ...581..817F},
BLAST \citep{2001dmsi.conf...59D},
% cf also 2004SPIE.5498...42D - but only abstract
BOLOCAM (LMT) and (CSO) \citep{2000ASPC..217..115M},
MAMBO \citep{2004MNRAS.354..779G},
SCUBA \citep{1999MNRAS.308..527B}, and SCUBA2 \citep{2004SPIE.5498...63A}.

\begin{figure}%[here!]
        %\resizebox{\hsize}{!}{\includegraphics{figs/Cl2048.pdf}}
        \resizebox{\hsize}{!}{\includegraphics{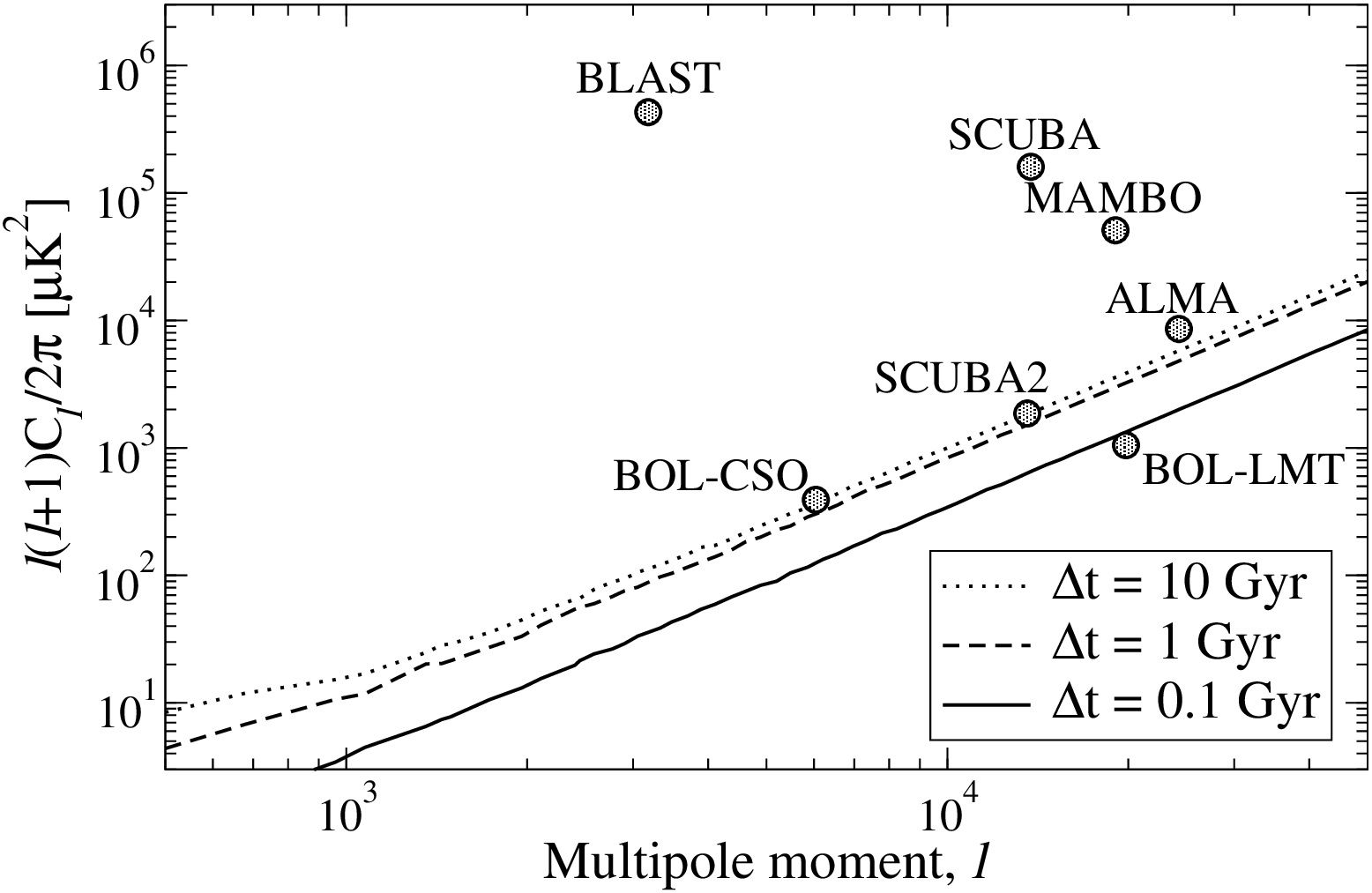}} % Cl2048_2.pdf
        \caption{Dust power spectrum as a function of multipole moment, $\ell$,
	for different dust lifetimes.
	%The left graph has logarithmic scales and
	%covers the multipole region $500\leq \ell \leq 500,000$.
	%The right graph has linear scales and covers the multipole region $500\leq \ell \leq 20,000$.
	Estimated detection limits for different instruments are also included as
	%dotted lines with $\ell \sim 1/FWHM$ marked with a small circle.}
	small circles placed at $\ell \sim 1/FWHM$.}
        %for a map 45'$\times$45' and resolution 30".}
        \label{fig:Cl2048}
\end{figure}

Using Eqs.~\ref{eq:ErrTot} and \ref{eq:Eerr} we have estimated the sensibilities
of these detectors. 
The result is presented in Table~\ref{tb:det}. % and Fig \ref{fig:Cl2048}.
Since all of these instruments operate
on a small patch in the sky we use $f_{sky}=f_{cut}=10\cdot FOV$, where $FOV$ is
the field of view of the instrument. The integration
time was set to one hour and
the noise per second, $s_X$, was calculated as $s_X=NEFD/\sqrt{N_{det}}$,
where $N_{det}$ is the number of detectors and $NEFD$ is the noise equivalent flux density.
The error was evaluated at the multipole moment $\ell \sim 1/FWHM$ 
(for BOLOCAM(LMT), $\ell$ was set to 20,000),
and we used a bin-size of $L=\ell$.
Note that ALMA is an interferometer and thus 
 $E_{instr} = f_{sky}\frac{4\pi s_X^2}{t_{obs}} \cdot \frac{\ell(\ell+1)}{2\pi}$
 %the exponential factor in Eq.~\ref{eq:ErrTot} drops out
Furthermore, the SCUBA2 array needs to be renormalized such that
$NEFD = NEFD\times \sqrt{N_{det}/(FOV/(\pi/4 \cdot (FWHM/60)^2))}$.
The resulting sensitivities $\sigma^{instr}_\ell$
%\begin{equation}
%	\sigma^{instr}_\ell = \sqrt{\frac{2}{(2\ell+1)f_{cut}L}} \times E_{instrument},
%\end{equation}
can be compared with the dust signal, as plotted in Fig.~\ref{fig:Cl2048}.
As can be seen, BLAST, SCUBA and MAMBO are unable to detect
the dust signal. However, BOLOCAM(LMT), ALMA, SCUBA2, and BOLOCAM(CSO) have good chances of detecting
the radiation from the first dust. 
We also note that the curves are almost parabolic. In fact, for
$1,000\lesssim\ell\lesssim 100,000$ the curves can be fitted within $\sim 10\%$ as:
\begin{equation}
\begin{array}{cccccr}
\ell(\ell+1)C_\ell^{Dust}/2\pi  & \approx & 2.13\times 10^{-5} \times \ell^{1.92},&\; \Delta t&=&10\textrm{ Gyr}, \\
\ell(\ell+1)C_\ell^{Dust}/2\pi   & \approx & 1.37\times 10^{-5} \times \ell^{1.95},& \; \Delta t&=&1\textrm{ Gyr},\\
\ell(\ell+1)C_\ell^{Dust}/2\pi & \approx & 4.02\times 10^{-5} \times \ell^{1.98},& \; \Delta t&=&0.1\textrm{ Gyr},
\end{array}
\end{equation}
in units of $\mu$K$^2_\textrm{CMB}$.
The dependency in $\ell$, which is slightly different from an uncorrelated
noise $\ell^2$ behavior, means that large-scale correlations cannot be neglected.
They mix differently at different epochs, depending on the dust lifetime parameter.

%First we compare with SCUBA measurements, see \cite{1999MNRAS.308..527B}, and
%find that at the mean $\ell=13081$ of SCUBA, the upper detection limit is
%$\left.\ell(\ell+1)C^{SCUBA}_\ell/2\pi\right|_{\ell=13081} \approx 55450$ $\mu$K$^2$.
%For this value of $\ell$, the dust has $\ell(\ell+1)C^{Dust}_\ell/2\pi \approx 1000$ $\mu$k$^2$
%which is hence a much too weak signal for SCUBA.

\begin{small}
\begin{table}[here!]
\caption{Sensitivities, $\sigma^{instr}_\ell$, for different (current and future) detectors.
	$NEFD$ = noise equivalent flux density, $\nu$ is the operating instrument frequency,
	$N_{det}$ = number of detectors, $FWHM$ = full width at half max, $FOV$ = field of view in
	units of arcmin$^2$, $\ell=1/FWHM$, $\sigma^{instr}_\ell$
	is the instrument sensitivity in units of mK$_{\textrm{CMB}}^2$.
	The instrument sensitivity was calculated with Eqs.~\ref{eq:ErrTot} \& \ref{eq:Eerr} using $t_{obs}=1$h,
	$f_{sky}=f_{cut}=10\cdot FOV$ and $L=\ell$.}
\setlength{\tabcolsep}{3pt}
\begin{tabular}{@{}lrrrrrrr@{}}
\hline
\hline
Instrument & $NEFD$             & \multicolumn{1}{c}{$\nu$} & $N_{det}$ & $FWHM$     & $FOV$   & \multicolumn{1}{c}{$\ell$}       & \multicolumn{1}{c}{$\sigma^{instr}_\ell$}    \\%& $\frac{\ell(\ell+1)}{2\pi}C_\ell^{D}$ \\
&$\left[\frac{\textrm{mJy}}{\sqrt{\textrm{Hz}}}\right]$ & [GHz] &           & [$"$] & [$'^2$] & $[10^3]$ &  [$10^3\mu$K$^2$] \\%& [mK] \\
\hline
SCUBA   &75	&353	&37	&13.8	&4.2	&14	&161	\\ %&1.3 \\
SCUBA2  &25	&353	&5120	&14.5	&50	&14	&1.8 	\\ %&1.3 \\
%MAMBO   &30-45	&250	&117	&10.7	&13	&19	&22-50	\\ %&2.8 \\
MAMBO   &45	&250	&117	&10.7	&13	&19	&50	\\ %&2.8 \\
BLAST   &239	&600	&43	&59	&85	&3.5	&424	\\ %&0.1 \\
\multicolumn{2}{@{}l}{BOLOCAM} &&&&&& \\
(CSO)   &40	&280	&144	&31	&50	&6.6	&0.38 	\\ %&0.19 \\
(LMT)   &3	&280	&144	&6	&3.1	&20	&1.0	\\ %&0.19 \\
ALMA    &1.5	&353	&1	&13.8/2	&0.085	&24	&8.0	\\ %&2867 \\ 
%ALMA    &1.5	&353	&64	&0.4	&0.02	&20	&2073	\\ %&2867 \\ FWHM=14" => 2e-3 !
\hline
\end{tabular}
\label{tb:det}
\end{table}

%\begin{table}[here!]
%\begin{tabular}{llllllll}
%NEFD             & $\nu$ & $N_{det}$ & FWHM     & FOV          & $\ell$    & $E^{instr}_\ell$                & $\frac{\ell(\ell+1)}{2\pi}C_\ell^{Dust}$ \\
%%$\frac{\textrm{mJy}}{\sqrt{\textrm{Hz}}}$ & GHz &           & " & ['$^2$] & $10^3$    & 10$^3$$\mu$K & 10$^3$$\mu$K \\
%\end{tabular}
%\caption{Sensitivities for different (current and future) detectors.}
%\end{table}
\end{small}

%First we compare with SCUBA measurements, see \cite{1999MNRAS.308..527B}, and
%find that $\ell(\ell+1)C^{Dust}_\ell/2\pi$ at the mean $\ell=13081$ of SCUBA is
%$\sim 1000$ $\mu$K$^2$, see figure \ref{fig:Cl45}, which is much less than the upper limit of 
%$\left.\ell(\ell+1)C^{SCUBA}_\ell/2\pi\right|_{\ell=13081} \approx 55450$ $\mu$K$^2$ (upper-limit).
%This means that the dust signal is too weak to detect with SCUBA.

%Other planned instruments that might be of interest are
%ALMA, \cite{2003SPIE.4837..110W},
%%FIRAS II, \cite{2002ApJ...581..817F},
%BLAST, \cite{2001dmsi.conf...59D},
%BOLOCAM(LMT), \cite{2000ASPC..217..115M},
%MAMBO, \cite{2004MNRAS.354..779G} and
%SCUBA2.
%ALMA will have a resolution of $\lesssim 0".02$ and a sensitivity of 1.9 mJy/Hz$^{1/2}$
%$\sim 10^{-13}$W/m$^2$.
%BLAST will have a resolution of $\sim$0'.5 and a sensitivity of 40 mJy/Hz$^{1/2}$.
%BOLOCAM(LMT) will have a resolution of 8" and a sensitivity of 2 mJy/Hz$^{1/2}$.
%MAMBO has a resolution of 10.7" and a sensitivity of 30-45 mJy/Hz$^{1/2}$.
%SCUBA2 will have a resolution of 14".7 and a sensitivity of 50 mJy/Hz$^{1/2}$.
%a factor two better than SCUBA's. Unfortunately Herschel does not cover the  
% submillimeter frequencies of interest.

\section{Conclusions}
It seems that it is possible to detect the dust from the first generation of
stars with the Planck satellite on small angular scales ($\ell \gtrsim 1000$).
However, the detectability depends on the actual distribution of dust in the
early universe, and also, to a large extent, on the dust lifetime.

The results
are parametrized so that changing the frequency and the fraction of produced
metals that become dust is only a matter of scaling the figures:
$C_\ell \propto (\nu/353\textrm{ GHz})^4$ and
$C_\ell \propto (f_d/0.3)^2$.
The spectrum of the early dust is compared to that of the primary CMB
anisotropies, as well as that of the local dust. % and found to have a unique signature.
The unique spectral signature of the early dust will help in disentangling it from the CMB and the
different foregrounds (local dust and extragalactic far infrared background).

The spatial signature of the early dust is found to have $C_\ell \approx $ constant
$\approx 10^{-5}-10^{-4}$ $\mu$K$^2_\textrm{CMB}$ depending on the dust lifetime, $\Delta t$.
Obviously, other signals that are correlated with the
structures will also show a similar behavior in the power spectrum.
Notably, the near infrared background from primordial galaxies could
be correlated with the early dust.
%In this case, the spectral
%signature of the dust can be used to distinguish between the two
%signals.

%Furthermore, 
The next generation of submillimeter instruments will be adequate to measure
these early dust anisotropies at very small angular scales ($\ell \gtrsim 2000$).
Our estimation shows that BOLOCAM, SCUBA2 and ALMA have a good prospect
of finding the early dust.
However, for these instruments, more detailed simulations are required in order
to obtain a realistic DM and baryon distribution. 
A DM simulation on a smaller box, maybe $L=50h^{-1}$ for PLANCK and smaller
still for ALMA, would improve the results on the relevant
angular scales, $\ell\gtrsim 1000$. This also means that the particles are
smaller, giving a better level of detail.
Furthermore, the distribution of dust relative to the DM can also
be improved and it is even possible to include some semi-analytical
results from the galaxy simulations in GalICS.

\begin{acknowledgements}
Finally, we wish to express our gratitude towards Eric Hivon who wrote
the code to analyze the simulated signal and thus produce the dust power spectrum.
\end{acknowledgements}
%In order to improve on these results, a more 

%A brief discussion of the model, its limits and ways to improve it  
%would be welcome. Clearly, the mass resolution may be an issue (to be  
%checked). The way dust is distributed within DM clumps (or  the central  
%zones of DM clumps, or a little beyond DM clumps) can be studied  
%further...

%As we have shown, the dust will not be a detectable contaminant of the
%Planck mission. However, other future instruments with higher sensitivity
%and angular resolution might detect it...

%--------------------------------------------------------------------
\bibliographystyle{aa} % style aa.bst
%--------------------------------------------------------------------

\bibliography{bibtex}        % bibtex.bib is the name of our database

\end{document}